\documentclass[11pt]{article}

\usepackage[final]{acl}

\usepackage{times}
\usepackage{latexsym}
\usepackage{booktabs}
\usepackage{tikz}
\usetikzlibrary{arrows.meta,positioning}
\usepackage{multirow}
\usepackage{graphicx}
\usepackage{tikz}
\usetikzlibrary{arrows.meta}
\usepackage{amsmath}
\usepackage{array}
\usepackage{tabularx}
\usepackage{listings}
\usepackage{subcaption}
\usepackage{enumitem}
\usepackage{hyperref}
\usepackage{fontawesome5}

\usepackage[T1]{fontenc}

\usepackage[utf8]{inputenc}

\usepackage{microtype}

\usepackage{inconsolata}

\title{Multi-View Decompilation for LLM-Based Malware Classification}

\author{
\begin{tabular}{cc}
\begin{tabular}{c}
Bercan Efe \\
Independent Researcher \\
\texttt{befeturkmen@gmail.com}
\end{tabular}
&
\begin{tabular}{c}
Vyas Raina \\
SPARK \\
\texttt{vyas@sparkairesearch.com}
\end{tabular}
\\[1.0em]
\multicolumn{2}{c}{
\small
\href{https://github.com/B3f3/Compiled-C-Malware}{
\faGithub\ \texttt{github.com/B3f3/Compiled-C-Malware}
}
}
\end{tabular}
}

\begin{document}
\maketitle

\begin{abstract}
Malware analysts often inspect compiled binaries through decompiled pseudo-C, when source code is unavailable. Recent work suggests that large language models (LLMs) can assist this process by classifying decompiled code as benign or malicious, but existing pipelines typically rely on a single decompiler view. We argue that this assumption is fragile: decompilers are lossy heuristic tools, and different decompilers can expose different artefacts of the same binary. We curate a benchmark of benign utilities and malicious programs spanning a range of threat behaviors. Each sample is compiled and decompiled with both Ghidra and RetDec, yielding matched pseudo-C views. Across a range of LLMs from major model families, we find that providing both decompiler views improves malicious-class $F_1$, mainly by increasing recall on malicious samples. Agreement analyses further show that Ghidra and RetDec make partially different errors, supporting the view that decompiler outputs provide complementary evidence. Our results suggest that multi-decompiler prompting is a simple, training-free way to improve LLM-based malware triage in practical settings.
\end{abstract}

\section{Introduction}

Software is often distributed as compiled binaries rather than source
code. This is common across commercial software, games, security tools,
and enterprise applications, where source code is withheld to protect
intellectual property or simplify deployment. However, binaries are also
harder to audit: a program may appear to provide benign functionality
while also containing malicious behavior such as credential theft,
process injection, file encryption, or command-and-control communication.
At the scale faced by malware triage teams, manually inspecting every
suspicious binary is infeasible.

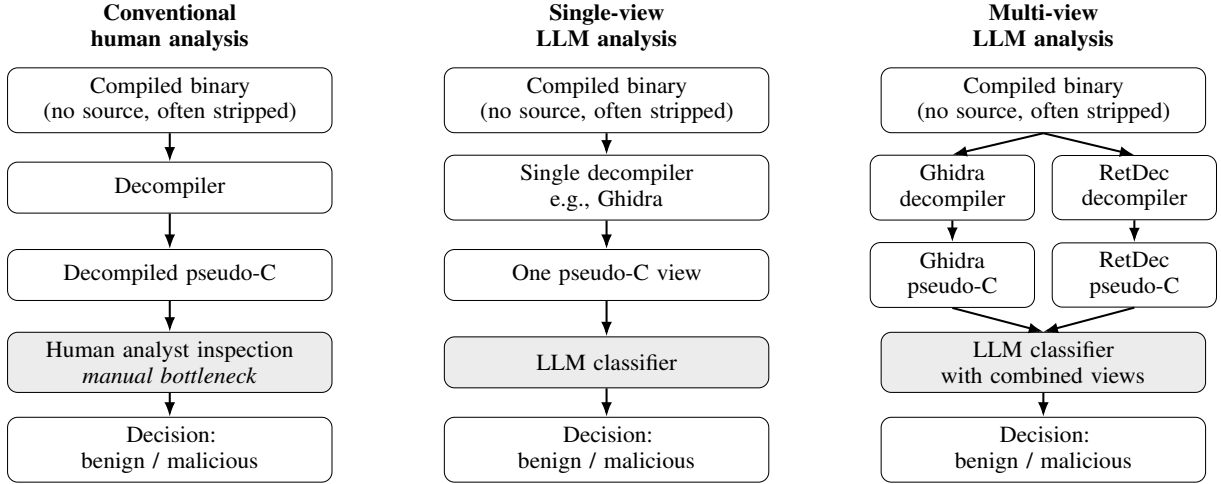
\begin{figure*}[t]
\centering
\resizebox{\textwidth}{!}{%
\begin{tikzpicture}[
  arr/.style={-{Latex[length=2mm]}, thick},
  title/.style={font=\bfseries\small, align=center, text width=4.2cm},
  stage/.style={
    draw, rounded corners, align=center, font=\small,
    text width=4.2cm, minimum height=7mm
  },
  smallstage/.style={
    draw, rounded corners, align=center, font=\small,
    text width=2.0cm, minimum height=7mm
  },
  bottleneck/.style={
    draw, rounded corners, align=center, font=\small,
    text width=4.2cm, minimum height=7mm, fill=gray!15
  },
  llm/.style={
    draw, rounded corners, align=center, font=\small,
    text width=4.2cm, minimum height=7mm, fill=gray!15
  }
]

\def\xone{0}
\def\xtwo{6}
\def\xthree{12}

\def\ytitle{0}
\def\ybin{-1.0}
\def\ydec{-2.2}
\def\ypc{-3.4}
\def\yagent{-4.6}
\def\yout{-5.8}

\node[title] at (\xone,\ytitle) (title1)
  {Conventional\\human analysis};
\node[stage] at (\xone,\ybin) (bin1)
  {Compiled binary\\(no source, often stripped)};
\node[stage] at (\xone,\ydec) (dec1)
  {Decompiler};
\node[stage] at (\xone,\ypc) (pc1)
  {Decompiled pseudo-C};
\node[bottleneck] at (\xone,\yagent) (hum1)
  {Human analyst inspection\\\emph{manual bottleneck}};
\node[stage] at (\xone,\yout) (out1)
  {Decision:\\benign / malicious};

\draw[arr] (bin1) -- (dec1);
\draw[arr] (dec1) -- (pc1);
\draw[arr] (pc1) -- (hum1);
\draw[arr] (hum1) -- (out1);

\node[title] at (\xtwo,\ytitle) (title2)
  {Single-view\\LLM analysis};
\node[stage] at (\xtwo,\ybin) (bin2)
  {Compiled binary\\(no source, often stripped)};
\node[stage] at (\xtwo,\ydec) (dec2)
  {Single decompiler\\e.g., Ghidra};
\node[stage] at (\xtwo,\ypc) (pc2)
  {One pseudo-C view};
\node[llm] at (\xtwo,\yagent) (llm2)
  {LLM classifier};
\node[stage] at (\xtwo,\yout) (out2)
  {Decision:\\benign / malicious};

\draw[arr] (bin2) -- (dec2);
\draw[arr] (dec2) -- (pc2);
\draw[arr] (pc2) -- (llm2);
\draw[arr] (llm2) -- (out2);

\node[title] at (\xthree,\ytitle) (title3)
  {Multi-view\\LLM analysis};
\node[stage] at (\xthree,\ybin) (bin3)
  {Compiled binary\\(no source, often stripped)};

\node[smallstage] at (10.75,\ydec) (ghdec)
  {Ghidra\\decompiler};
\node[smallstage] at (13.25,\ydec) (rtdec)
  {RetDec\\decompiler};

\node[smallstage] at (10.75,\ypc) (ghpc)
  {Ghidra\\pseudo-C};
\node[smallstage] at (13.25,\ypc) (rtpc)
  {RetDec\\pseudo-C};

\node[llm] at (\xthree,\yagent) (llm3)
  {LLM classifier\\with combined views};
\node[stage] at (\xthree,\yout) (out3)
  {Decision:\\benign / malicious};

\draw[arr] (bin3.south) -- (ghdec.north);
\draw[arr] (bin3.south) -- (rtdec.north);
\draw[arr] (ghdec) -- (ghpc);
\draw[arr] (rtdec) -- (rtpc);
\draw[arr] (ghpc.south) -- (llm3.north);
\draw[arr] (rtpc.south) -- (llm3.north);
\draw[arr] (llm3) -- (out3);

\end{tikzpicture}%
}
\caption{
Three workflows for binary malware classification. Conventional reverse
engineering relies on a human analyst to inspect decompiled pseudo-C.
A single-view LLM pipeline replaces this manual stage with an LLM, but
still depends on one decompiler's representation. This work studies a
multi-view alternative in which outputs from multiple decompilers provide
complementary views of the same binary to the LLM.
}
\label{fig:pipeline}
\end{figure*}

In the absence of source code, analysts typically rely on
\emph{decompilers} \citep{ghidra, idapro}. A decompiler attempts to
recover an approximate, human-readable C-like representation of a program
from a compiled binary \citep{cifuentes1994reverse, cao2024decompilers}. Reverse engineers then inspect this
decompiled pseudo-code to reason about the program's behavior and decide
whether it is benign or malicious. Although this workflow is standard,
the inspection stage remains a costly human bottleneck: it requires
specialist expertise, is time-consuming, and does not scale easily.
Figure~\ref{fig:pipeline} illustrates this conventional workflow and
contrasts it with the LLM-based single-view and multi-view pipelines
studied in this paper.

The code-understanding capabilities of large language models (LLMs) make
them a natural candidate for automating part of this inspection process
\citep{zhang2024cybersecurity, xu2024llmcyber}.
Recent work suggests that LLMs can interpret decompiled code and reason
about program behavior
\citep{pordanesh2024gpt4, fang2024llmcode, manuel2024debinvul}. A
straightforward pipeline is therefore to decompile a binary, pass the
pseudo-code to an LLM, and ask the model to classify the sample as benign
or malicious. Such a system could support first-pass triage, reduce
analyst workload, and surface suspicious samples for human review.

However, most LLM-based decompilation pipelines rely on a single
decompiler, often treating its output as the canonical representation of
the program \citep{chawla2026decompilation, manuel2024debinvul}. This
assumption is fragile. Decompilation is a lossy and heuristic process:
different tools make different choices about control-flow recovery,
variable naming, type inference, and expression simplification
\citep{dramko2024taxonomy, lee2011tie, schwartz2013native}. As a
result, each decompiler exposes a different \emph{view} of the same
binary. These views can contain tool-specific artefacts that are not
semantically neutral for an LLM \citep{cao2024decompilers}. A behavior
obscured in one decompiler's output may be more visible in another, while
a noisy representation may also mislead the model.

This motivates a simple hypothesis: different decompilers can provide
complementary evidence to an LLM. If the errors induced by one
decompiler's representation are not perfectly correlated with those
induced by another, then combining multiple decompiled views may improve
classification. This is attractive in practice because it can be built on
top of existing open-source decompilers and off-the-shelf LLMs, without
requiring model training or access to source code.

We study this hypothesis empirically. We construct a balanced benchmark
of benign and malicious C programs, compile each program into a binary,
and decompile every binary with two widely used free and open-source
decompilers: Ghidra and RetDec \citep{ghidra, retdec, cao2022neurdp}. 
The benign samples are curated to reflect
common utility behaviors, including parsing, cryptography, monitoring,
and networking. The malicious samples cover a range of malware
behaviors, including botnets, worms, trojans, rootkits, backdoors, and
keylogging. This setup lets us evaluate both how well off-the-shelf LLMs
classify decompiled binaries and whether multiple decompiler views
provide useful complementary information.

Across a panel of LLMs, we find that decompiler choice substantially
affects classification behavior, and that combining decompiler views
improves performance in most settings across model families. The gains
are particularly important for smaller models, where multi-view inputs
improve recall. We also find that disagreements between decompiler views
are informative: when single-view predictions differ, the additional
context from multiple views often helps the model recover the correct
decision. Our main contributions are:
\begin{itemize}[leftmargin=*, itemsep=1pt, topsep=2pt]
    \item A curated benchmark for LLM-based malware classification from
    decompiled binaries, with matched Ghidra and RetDec outputs.

    \item A multi-view formulation for comparing single- and
    multi-decompiler LLM pipelines.

    \item An empirical study showing that decompiler views are
    complementary and that combining them improves performance in most
    settings.
\end{itemize}

\section{Related Work}
\label{sec:related}
 
\paragraph{LLMs for malware detection.}
Large language models are increasingly used to analyse code for security
defects, including detecting vulnerabilities in source code across many
languages \citep{dozono2024secure,jiang2024cvd} and repairing buggy
programs \citep{xia2023apr}, with recent surveys mapping the breadth of
these security applications
\citep{zhang2024cybersecurity,xu2024llmcyber}. Malware detection is a
prominent instance of this trend. Beyond decompiled binaries, LLMs and
earlier transformer models have been applied across input
representations: detecting injected malicious functions in source
packages \citep{tsfaty2022malicious}, classifying Android applications
from multiple feature views
\citep{zhao2025apppoet,qian2025lamd,walton2025android}, and summarising
or deobfuscating real-world malware to support human analysts
\citep{fujii2024feasibility,patsakis2024deobfuscation,lu2024malsight},
with broader reviews surveying this fast-growing area
\citep{alkaraki2024malware}. The recurring finding is that LLMs carry
useful priors about malicious behavior from code-level inputs; our focus
is on surfacing those priors when the input is the noisy, decompiled C
recovered from a stripped binary rather than source code.
 
\paragraph{LLMs for decompiled-code analysis.}
Recent work has begun to use LLMs for analysing decompiled binaries.
\citet{pordanesh2024gpt4} study GPT-4's ability to reason about
reverse-engineered code, while \citet{manuel2024debinvul} benchmark LLMs
for vulnerability analysis over decompiled functions. Most closely
related to our work, \citet{chawla2026decompilation} propose a
decompilation-driven pipeline for malware detection with LLMs, addressing
the same end task that we study. These approaches typically operate on
the output of a single decompiler, often Ghidra. This makes the
decompiler's representation the sole view of the binary available to the
model. \citet{fang2024llmcode} show that LLM code analysis can be brittle
and sensitive to surface form, which is a particular concern when the
input is a noisy, tool-specific reconstruction rather than source code.
Our work keeps the decompiler--LLM pipeline, but studies whether multiple
decompiler views provide complementary evidence for malware
classification.
 
\paragraph{Decompiler variability.}
Decompiler outputs can differ substantially across tools.
\citet{cao2024decompilers} evaluate the effectiveness of current
decompilers, and \citet{dramko2024taxonomy} characterise fidelity issues
in recovered C, including errors in type recovery, naming, and
control-flow reconstruction. This difficulty is intrinsic to the task:
decompilation must reconstruct high-level types and control flow that
compilation discards, a problem studied since early work on reverse
compilation and type recovery
\citep{cifuentes1994reverse,lee2011tie,schwartz2013native}. This line of
work shows that decompiled pseudo-code is not a canonical representation
of the original program: each tool makes heuristic choices that can
change what is exposed, obscured, or distorted. Prior work mainly treats
this variability as a challenge for human reverse engineers. We instead
ask whether such variability can also be useful for LLM-based analysis,
by providing multiple imperfect but complementary views of the same
binary.
 
\paragraph{Combining multiple views.}
Combining diverse predictions is a standard way to improve robustness in
machine learning, including through classical model ensembles and, more
recently, LLM methods that aggregate over multiple reasoning paths
\citep{wang2022selfconsistency,li2024moreagents} or over the outputs of
several prompts, agents, or models
\citep{du2023multiagentdebate,wang2024moa,chen2025llmensemble}. These
methods usually vary the model-side computation while keeping the input
fixed. Our work considers a different axis of diversity: the input
representation itself. By decompiling the same binary with multiple
tools, we obtain distinct pseudo-code views whose errors and artefacts
need not be correlated. We then study whether these views can be combined
to improve LLM-based malware classification. To our knowledge, prior work
on LLM-based malware detection from decompiled binaries has not
systematically evaluated multi-decompiler aggregation.

\section{Multi-View Decompiler Classification}
\label{sec:method}

\paragraph{From binaries to decompiled views.}
Our setting follows the static reverse-engineering pipeline introduced in
Figure~\ref{fig:pipeline}. Given a compiled binary, a decompiler attempts
to recover an approximate C-like representation that can be inspected by
a human analyst or, in our case, by an LLM. Decompilation is more
informative than raw disassembly for code understanding, but it is also
lossy and heuristic: tools must reconstruct control flow, infer types,
recover variables, and assign names without access to the original source~\citep{cifuentes1994reverse, schwartz2013native, lee2011tie}. These
choices are not uniquely determined, so the same binary can yield
different pseudo-code depending on the decompiler.

This tool dependence is central to our study. Prior work shows that
current decompilers differ in the readability and semantic fidelity of
their outputs~\citep{cao2024decompilers, dramko2024taxonomy}. In
practice, Ghidra and RetDec often expose different surface forms of the
same program, including different naming conventions, type recovery, cast
structure, and control-flow reconstruction. We treat these outputs as
distinct \emph{views} of the same binary. Rather than assuming that one
decompiler provides the canonical representation, we ask whether multiple
views provide complementary evidence for LLM-based malware
classification.

\paragraph{Task formulation.}
Let $b$ denote a compiled binary with ground-truth label
$y(b)\in\{0,1\}$, where $1$ denotes malicious and $0$ denotes benign. We
decompile each binary with two decompilers,
$\mathcal{D}=\{\mathrm{G},\mathrm{R}\}$, corresponding to Ghidra and
RetDec. Each decompiler $\delta_d$ produces a pseudo-C view
\[
    c_d = \delta_d(b), \qquad d \in \{\mathrm{G},\mathrm{R}\}.
\]
The task is to predict $y(b)$ from one or more decompiled views. We use
Ghidra and RetDec, two of the most widely used decompilers~\citep{cao2022neurdp, chawla2026decompilation}, because both are free, open source,
scriptable, and suitable for reproducible batch processing, while still
relying on different decompilation pipelines.

\paragraph{LLM classifier.} We use an LLM as a binary classifier over decompiled code. A prompt
template $\pi$ wraps one or more pseudo-C views with task instructions and
asks the model to reason about the program before returning a final label
in $\{0,1\}$. The reasoning step is included because, in pilot
experiments, it improved classification reliability~\footnote{The prompt templates are given in Appendix~\ref{app:prompts}}. The final decision
is read directly from the model's binary output; we do not use model
probabilities or apply a threshold.

\paragraph{Single-decompiler prediction.}
The single-view setting classifies each decompiler output independently.
For decompiler $d$, the prediction is
\[
    \hat{y}_{d}
    =
    f\!\left(\pi_{\mathrm{single}}(c_d)\right),
    \qquad d \in \{\mathrm{G},\mathrm{R}\},
\]
where $f$ is the LLM and $\pi_{\mathrm{single}}$ is the prompt used for a
single pseudo-C view. This setting tests how much classification
performance depends on the choice of decompiler.

\paragraph{Multi-view prediction.}
The multi-view setting presents both decompiler outputs to the model in a
single prompt:
\[
    \hat{y}_{\mathrm{GR}}
    =
    f\!\left(\pi_{\mathrm{multi}}(c_{\mathrm{G}}, c_{\mathrm{R}})\right).
\]
The prompt tells the model that the two inputs are complementary
decompiled views of the same binary and asks it to classify the binary as
malicious if either view provides convincing evidence of malicious
behavior. This setting tests whether an LLM can benefit from seeing two
tool-specific reconstructions of the same program.

\paragraph{Disagreement-triggered consensus.}
We also consider a simple consensus rule that uses the multi-view prompt
only when the two single-view predictions disagree:
\[
  \hat{y}_{\mathrm{cons}} =
  \begin{cases}
    \hat{y}_{\mathrm{G}} & \text{if } \hat{y}_{\mathrm{G}} = \hat{y}_{\mathrm{R}},\\[2pt]
    \hat{y}_{\mathrm{GR}} & \text{otherwise.}
  \end{cases}
\]
The motivation is that agreement between decompiler views provides a
simple signal of confidence, while disagreement identifies cases where a
single decompiler choice would change the decision. In deployment, this
rule avoids running the longer multi-view prompt on every sample, but it
still requires the two single-view LLM calls to be made first. It
therefore trades additional inference cost on contested examples for a
more targeted use of combined decompiler context.

\section{Dataset Creation}
\label{sec:dataset}

\paragraph{Benchmark design.}
We construct a balanced benchmark of 100 C programs, with 50 benign and
50 malicious samples. Each program is compiled into a binary object and
decompiled with both Ghidra and RetDec, producing matched decompiler
views for every sample. This matched design ensures that all samples pass
through the same compilation and decompilation pipeline, so differences
in the pseudo-C presented to the LLM reflect decompiler-specific
behavior rather than differences in sample processing.

The benchmark is designed to reflect the ambiguity encountered in
practical malware triage. We intentionally construct the benign corpus
around realistic utility behaviors such as file processing, parsing,
cryptography, process and system monitoring, and networking. These
behaviors overlap with low-level idioms that also appear in malware, so
the classifier cannot rely on superficial cues such as the presence of
sockets, encryption, or filesystem access. The malicious corpus is
curated to cover a range of threat behaviors seen in security analysis,
including botnets, worms, banking trojans, rootkits, backdoors, and
keylogging. Table~\ref{tab:dataset-composition} summarises the dataset
composition.

\paragraph{Compilation and decompilation.}
All samples are compiled on Linux using a single \texttt{gcc}
configuration, producing 64-bit x86 ELF relocatable object files
(\texttt{.o}). We use \texttt{-O2} optimisation and apply the same
compiler settings uniformly across classes. We also use
\texttt{-fno-stack-protector} to avoid stack-canary instrumentation that
can obscure program logic in the decompiled output, and \texttt{-s} to
strip symbol information, reflecting the reduced symbolic information
available in many binaries encountered during analysis. Each compiled object is decompiled with Ghidra and RetDec.

\begin{table}[t]
\centering
\small
\setlength{\tabcolsep}{4pt}
\begin{tabular}{@{}llr@{}}
\toprule
\multicolumn{3}{@{}l}{\textbf{Benign} (50 samples)} \\
\midrule
Domain & Example programs & $n$ \\
\midrule
Core utilities   & \texttt{cat}, \texttt{grep}, \texttt{wc}        & 10 \\
Archive/parsing  & \texttt{base64}, \texttt{json\_tok}             & 10 \\
Cryptography     & \texttt{aes128}, \texttt{sha256}, \texttt{hmac} & 10 \\
Monitoring       & \texttt{proclist}, \texttt{meminfo}             & 10 \\
Networking       & \texttt{http\_get}, \texttt{tcp\_server}        & 10 \\
\midrule
\multicolumn{3}{@{}l}{\textbf{Malicious} (50 samples)} \\
\midrule
Family & behavior type & $n$ \\
\midrule
Mirai          & IoT botnet        & 13 \\
MyDoom         & Email worm        & 13 \\
Hellbot        & Botnet            & 8  \\
Dexter POS     & PoS malware       & 5  \\
Carberp        & Banking trojan    & 3  \\
Remhead        & Rootkit           & 2  \\
x0r USB Worm   & USB worm          & 2  \\
Keylogger      & Keylogger         & 1  \\
Minipig        & Backdoor/RAT      & 1  \\
Rovnix         & Bootkit           & 1  \\
Rubilyn        & Rootkit           & 1  \\
\bottomrule
\end{tabular}
\caption{Dataset composition. Benign programs are original utilities
spread evenly across five functional domains; malicious programs are
grouped by malware family and behavior type.}
\label{tab:dataset-composition}
\end{table}

\section{Experiments}
\label{sec:experiments}

\subsection{Experimental Setup}

\paragraph{Models.}
We evaluate a cost-conscious panel of instruction-tuned LLMs spanning
major model families and both proprietary and open-weight systems:
Google's \texttt{gemini-2.5-flash-lite} \citep{gemini25}, OpenAI's
\texttt{GPT-5.4-mini} \citep{gpt54mini}, Anthropic's
\texttt{Claude Haiku 4.5} \citep{haiku45}, Alibaba's open-weight
mixture-of-experts model \texttt{Qwen3-35B-A3B} \citep{qwen3}, and
Meta's open-weight \texttt{Llama-3.3-70B-Instruct} \citep{llama3}.
We focus on smaller or lower-cost variants of leading model families
because industrial malware triage may require running inference over
large volumes of binaries, making routine use of the largest frontier
models economically unattractive. This gives a more realistic deployment
setting: models must be capable enough to reason about decompiled code,
but cheap enough to run across many samples and decompiler views.

All models are queried at temperature $0.6$. Prompts ask the model to
give a brief rationale before returning structured JSON with a binary
decision, where $1$ denotes malicious and $0$ benign. Full prompt templates for every setting are listed in Appendix~\ref{app:prompts}. 
Each sample-setting pair is evaluated across five independent runs; we report mean performance and run-to-run standard deviation.

\paragraph{Data, settings, and metrics.}
We evaluate on the benchmark from Section~\ref{sec:dataset}. All 100
samples have both Ghidra and RetDec outputs, so each model is evaluated
on the full benchmark under single-decompiler, multi-view, and
disagreement-triggered consensus settings. We report accuracy, precision,
recall, and $F_1$ for the malicious class, where
$\mathrm{Precision}=\mathrm{TP}/(\mathrm{TP}+\mathrm{FP})$,
$\mathrm{Recall}=\mathrm{TP}/(\mathrm{TP}+\mathrm{FN})$, and
$F_1=2PR/(P+R)$. Since false negatives are especially costly in malware
triage, we emphasise recall and $F_1$ in the analysis.

\subsection{Results}

\paragraph{Main results.}

\begin{table*}[htb!]
\centering
\small
\setlength{\tabcolsep}{8pt}
\renewcommand{\arraystretch}{1.1}
\begin{tabular}{lccccc}
\toprule
\textbf{Setting} &
\texttt{Qwen3} &
\texttt{Llama} &
\texttt{Haiku} &
\texttt{Flash-Lite} &
\texttt{GPT-mini} \\
\midrule
Ghidra only &
$\mathbf{79.5}_{\pm 2.7}$ &
$75.0_{\pm 3.3}$ &
$88.9_{\pm 1.2}$ &
$75.0_{\pm 2.4}$ &
$71.8_{\pm 2.2}$ \\

RetDec only &
$66.7_{\pm 2.1}$ &
$81.0_{\pm 2.4}$ &
$88.9_{\pm 1.9}$ &
$73.4_{\pm 3.1}$ &
$61.1_{\pm 2.8}$ \\
\midrule
Multi-view &
$78.0_{\pm 0.7}$ &
$\mathbf{88.9}_{ \pm 1.3}$ &
$\mathbf{92.5}_{\pm 0.9}$ &
$\mathbf{88.9}_{\pm 2.0}$ &
$\mathbf{79.5}_{\pm 1.6}$ \\

Consensus &
$75.2_{\pm 1.8}$ &
$80.3_{\pm 2.1}$ &
$91.5_{\pm 0.9}$ &
$80.6_{\pm 3.0}$ &
$73.1_{\pm 3.1}$ \\
\bottomrule
\end{tabular}%
\caption{Malicious-class $F_1$ (\%) by model and decompilation setting.
Subscripts denote standard deviation over five runs.
Best result per model is bolded.}
\label{tab:main-results}
\end{table*}

\begin{figure*}[htb!]
\centering

\begin{subfigure}{0.32\textwidth}
\centering
\includegraphics[width=\linewidth]{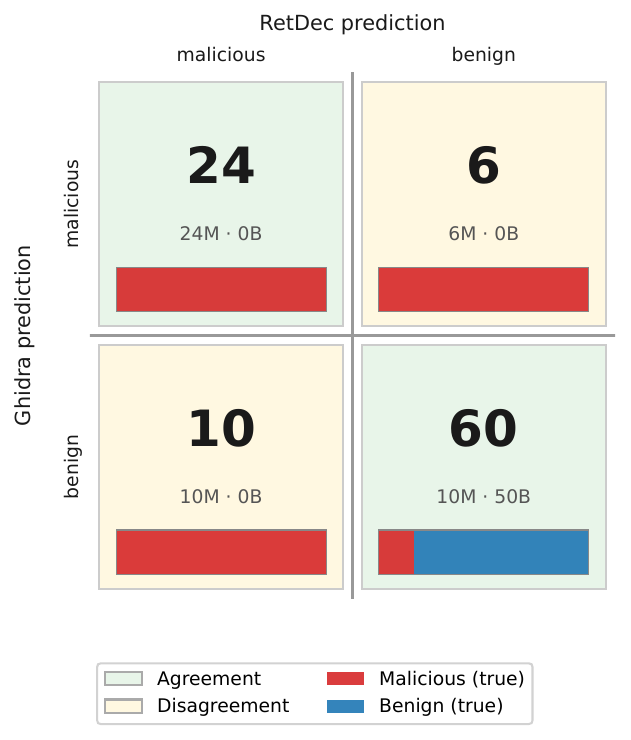}
\caption{\texttt{Llama}}
\end{subfigure}
\hfill
\begin{subfigure}{0.32\textwidth}
\centering
\includegraphics[width=\linewidth]{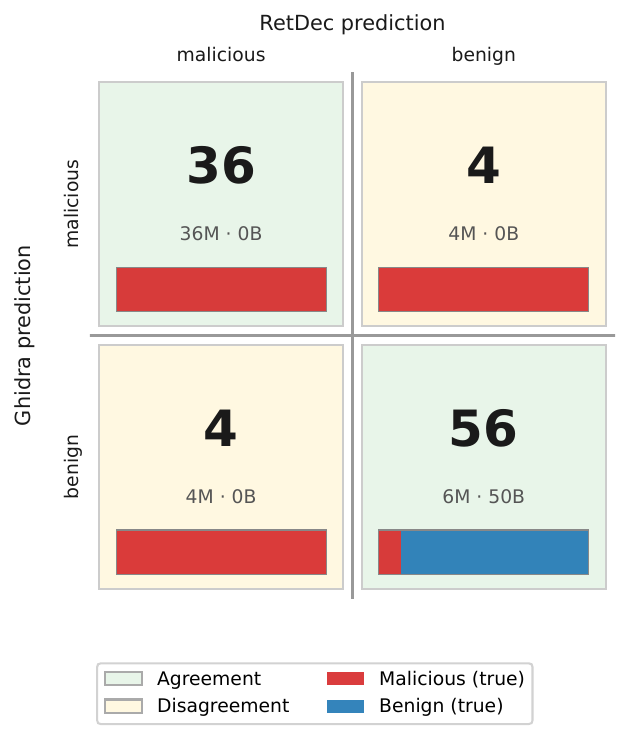}
\caption{\texttt{Haiku}}
\end{subfigure}
\hfill
\begin{subfigure}{0.32\textwidth}
\centering
\includegraphics[width=\linewidth]{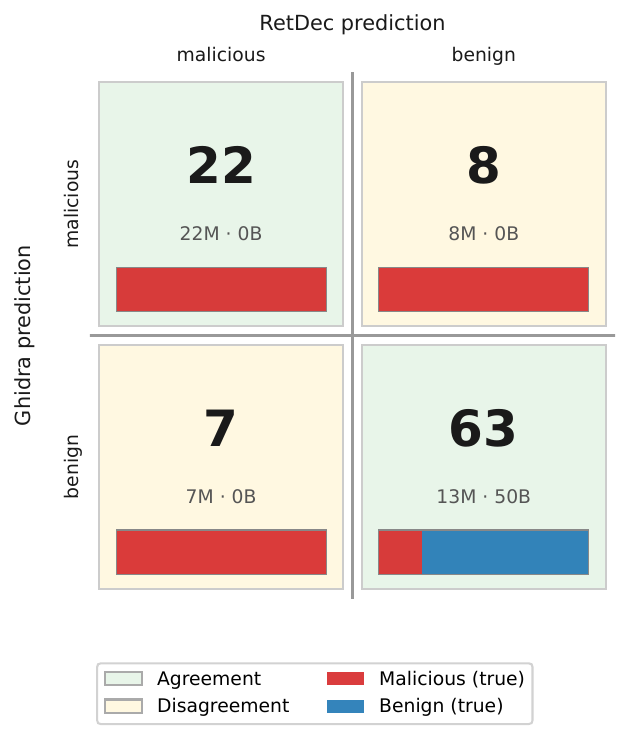}
\caption{\texttt{Flash-Lite}}
\end{subfigure}

\caption{Prediction agreement between Ghidra-only and RetDec-only
classifiers. Rows correspond to the Ghidra prediction and columns to the
RetDec prediction. Off-diagonal cells show samples where the two
decompiler views lead to different labels.}
\label{fig:agreement}
\end{figure*}

Table~\ref{tab:main-results} reports malicious-class $F_1$ for each
model and decompilation setting. All models obtain substantial
performance from decompiled pseudo-C, showing that off-the-shelf LLMs can
support malware classification in this setting. However, performance
depends strongly on both the model and the decompiler view. There is no
universally best single decompiler: Ghidra is stronger for Qwen and
\texttt{GPT-5.4-mini}, RetDec is stronger for \texttt{Llama-3.3-70B},
and the two are effectively tied for \texttt{Claude Haiku 4.5} and
\texttt{gemini-2.5-flash-lite}.

The multi-view setting is the strongest overall. It gives the best
$F_1$ for four of the five models, improving over the better single
decompiler by $+13.9$ for \texttt{gemini-2.5-flash-lite}, $+7.9$ for
\texttt{Llama-3.3-70B}, $+7.7$ for \texttt{GPT-5.4-mini}, and $+3.6$
for \texttt{Claude Haiku 4.5}. The exception is Qwen, where multi-view
classification is slightly below the best single view ($78.0$ vs.\
$79.5$). The disagreement-triggered consensus rule is competitive, but
it does not improve over direct multi-view prompting except for tying
\texttt{Claude Haiku 4.5}. Since direct multi-view classification uses a
single LLM call per sample, while consensus requires the two single-view
calls and sometimes an additional multi-view call, direct combination is
also the simpler option when the goal is final classification.

\paragraph{Decompiler complementarity.}
Figure~\ref{fig:agreement} shows prediction agreement matrices for the
single-decompiler settings, with all models reported in
Figure~\ref{fig:appendix-agreement}. Each subplot compares Ghidra-only
and RetDec-only predictions for one model; off-diagonal cells are samples
where changing only the decompiler view changes the predicted label.
These disagreements show that the tools are not interchangeable: each
can make different behaviours more or less visible through its choices
about control flow, types, casts, and naming.

This complementarity matters because the dominant failure mode is missed
malicious samples rather than false positives. Since precision is high,
$F_1$ gains mostly come from improved recall. When one view obscures the
relevant behaviour, the other may expose enough evidence to recover the
malicious label. Direct multi-view prompting also often beats the
consensus rule, suggesting that the gain is not limited to explicit
single-view disagreements: in some cases, neither view is sufficient
alone, but together they provide enough evidence for the model to make
the malicious call.

\section{Conclusions and Future Work}

We studied LLM-based malware classification from decompiled binaries and showed that decompiler choice materially affects performance. Single decompiler views often lead models to be conservative, missing malicious samples when the relevant behavior is obscured or rendered unclearly. Providing multiple decompiler views gives the model additional evidence, improving recall and therefore malicious-class $F_1$ in most settings; this is especially valuable for malware triage, where missed malicious samples are more costly than false positives. Our results suggest that decompiler diversity is a simple and practical way to improve off-the-shelf LLM analysis without model training. Future work should extend the benchmark with harder benign programs, broader malware behaviors, and emerging AI-era attack surfaces such as malicious or hallucinated dependencies in generated code. It should also explore richer combination strategies, including learned view selection, evidence aggregation, and cost-aware pipelines that decide when additional decompiler views are worth querying.

\section{Limitations}

\paragraph{Object files, not whole programs.}
We classify relocatable object files (\texttt{.o}), code fragments without a \texttt{main} entry point or full linkage. This mirrors how individual malware modules often appear, but a deployed analyst usually inspects a complete executable with surrounding context, so our findings may not transfer directly to whole-binary triage.
 
\paragraph{No obfuscation or anti-decompilation.}
We assume binaries decompile cleanly. In practice, authors sometimes deliberately obfuscate or pack code to resist decompilation, for instance to deter reverse engineering or code theft, and our samples are neither packed nor adversarially hardened (we retain part of the symbol table rather than scrambling or fully obscuring program structure). Heavily obfuscated or anti-decompilation-hardened binaries lie outside the setting we study and would likely degrade every approach we evaluate.
 
\paragraph{Number of decompilers.}
We study two decompilers, Ghidra and RetDec. Although these two are widely used, they do not cover other proprietary tools (e.g.\ IDA Pro, Binary Ninja), and our complementarity findings may change with other tools, or with more than two views combined.

\section{Ethics and Impact Statement}

\paragraph{Intended use and impact.}
We study the automated classification of binaries as malicious or benign,
which is a defensive task. Our aim is to help analysts, researchers, 
and security systems triage software more reliably, especially when 
the source code is unavailable and only decompiled output can be inspected. Our goal is detection rather than generating attacks: we do not develop novel malware, attack techniques, or evasion methods. 

\paragraph{Malicious samples.}
All but one of our malicious samples are real-world specimens of
well-documented families obtained from an established public repository \citep{thezoo}; the exception is the synthetic keylogger noted above,
which reimplements a publicly described technique and adds no new
capability. We introduce no new malware families, capabilities, or evasion
methods. Every program in the benchmark, benign and malicious alike, is
compiled from source and then statically decompiled inside an isolated,
network-disabled container; the malicious binaries are never executed.

\paragraph{Release and dual-use considerations.}
To support reproducibility, we share the decompiled artefacts and
evaluation code underlying our results for research purposes. Because the
malicious samples are either already-public specimens or a trivial
reimplementation of a long-documented technique, releasing their
decompiled representations does not provide capabilities beyond what the
community can already obtain, and we judge the reproducibility benefit to
outweigh this marginal risk. We nonetheless acknowledge the dual-use
nature of malware research: a determined adversary could attempt to craft
samples that evade an LLM-based classifier such as ours. Our released
material concerns detection rather than evasion and contains no new
offensive functionality. We follow the usage terms of the source
repositories and recommend that anyone reusing these artefacts do so only
in controlled, non-production environments for defensive research.


\bibliography{custom}

@misc{chawla2026decompilation,
  title        = {A Decompilation-Driven Framework for Malware Detection with Large Language Models},
  author       = {Chawla, Aniesh and Prasad, Udbhav},
  year         = {2026},
  eprint       = {2601.09035},
  archivePrefix = {arXiv},
  primaryClass = {cs.CR},
  url          = {https://arxiv.org/abs/2601.09035}
}

@misc{manuel2024debinvul,
  title        = {Enhancing Reverse Engineering: Investigating and Benchmarking Large Language Models for Vulnerability Analysis in Decompiled Binaries},
  author       = {Manuel, Dylan and Islam, Nafis Tanveer and Khoury, Joseph and Nunez, Ana and Bou-Harb, Elias and Najafirad, Peyman},
  year         = {2024},
  eprint       = {2411.04981},
  archivePrefix = {arXiv},
  primaryClass = {cs.CR},
  url          = {https://arxiv.org/abs/2411.04981}
}

@inproceedings{fang2024llmcode,
  title        = {Large Language Models for Code Analysis: Do {LLMs} Really Do Their Job?},
  author       = {Fang, Chongzhou and Miao, Ning and Srivastav, Shaurya and Liu, Jialin and Zhang, Ruoyu and Fang, Ruijie and Asmita and Tsang, Ryan and Nazari, Najmeh and Wang, Han and Homayoun, Houman},
  booktitle    = {33rd USENIX Security Symposium (USENIX Security 24)},
  pages        = {829--846},
  year         = {2024},
  isbn         = {978-1-939133-44-1},
  address      = {Philadelphia, PA},
  publisher    = {USENIX Association},
  url          = {https://www.usenix.org/conference/usenixsecurity24/presentation/fang}
}

@inproceedings{cao2024decompilers,
  title        = {Evaluating the Effectiveness of Decompilers},
  author       = {Cao, Ying and Zhang, Runze and Liang, Ruigang and Chen, Kai},
  booktitle    = {Proceedings of the 33rd ACM SIGSOFT International Symposium on Software Testing and Analysis (ISSTA 2024)},
  pages        = {491--502},
  year         = {2024},
  address      = {New York, NY, USA},
  publisher    = {Association for Computing Machinery},
  doi          = {10.1145/3650212.3652144},
  url          = {https://doi.org/10.1145/3650212.3652144}
}

@misc{pordanesh2024gpt4,
  title        = {Exploring the Efficacy of Large Language Models (GPT-4) in Binary Reverse Engineering},
  author       = {Pordanesh, Saman and Tan, Benjamin},
  year         = {2024},
  eprint       = {2406.06637},
  archivePrefix = {arXiv},
  url          = {https://arxiv.org/abs/2406.06637}
}

@misc{thezoo,
  author       = {ytisf and contributors},
  title        = {{theZoo}: A Live Malware Repository},
  year         = {2014},
  howpublished = {\url{https://github.com/ytisf/theZoo}},
  note         = {Accessed: June 2026}
}

@misc{ghidra,
  author       = {{National Security Agency}},
  title        = {Ghidra Software Reverse Engineering Framework},
  year         = {2019},
  howpublished = {\url{https://ghidra-sre.org}},
  note         = {Version 12.1, \url{https://github.com/NationalSecurityAgency/ghidra}}
}

@misc{retdec,
  author       = {{Avast Software}},
  title        = {{RetDec}: A Retargetable Machine-Code Decompiler},
  year         = {2024},
  howpublished = {\url{https://github.com/avast/retdec}}
}

@misc{gcc,
  author       = {{GCC Team}},
  title        = {{GNU Compiler Collection (GCC)}},
  year         = {2024},
  howpublished = {\url{https://gcc.gnu.org}},
  note         = {Version 13}
}

@phdthesis{cifuentes1994reverse,
  author = {Cifuentes, Cristina},
  title  = {Reverse Compilation Techniques},
  school = {Queensland University of Technology},
  year   = {1994}
}

@inproceedings{lee2011tie,
  author    = {Lee, JongHyup and Avgerinos, Thanassis and Brumley, David},
  title     = {{TIE}: Principled Reverse Engineering of Types in Binary Programs},
  booktitle = {Proceedings of the Network and Distributed System Security Symposium (NDSS)},
  year      = {2011}
}

@inproceedings{schwartz2013native,
  author    = {Schwartz, Edward J. and Lee, JongHyup and Woo, Maverick and Brumley, David},
  title     = {Native x86 Decompilation Using Semantics-Preserving Structural Analysis and Iterative Control-Flow Structuring},
  booktitle = {22nd USENIX Security Symposium (USENIX Security 13)},
  pages     = {353--368},
  year      = {2013},
  publisher = {USENIX Association}
}

@inproceedings{dramko2024taxonomy,
  author    = {Dramko, Luke and Lacomis, Jeremy and Schwartz, Edward J. and Vasilescu, Bogdan and Le Goues, Claire},
  title     = {A Taxonomy of {C} Decompiler Fidelity Issues},
  booktitle = {33rd USENIX Security Symposium (USENIX Security 24)},
  year      = {2024},
  publisher = {USENIX Association},
  url       = {https://www.usenix.org/conference/usenixsecurity24/presentation/dramko}
}

@misc{idapro,
  author       = {{Hex-Rays}},
  title        = {{IDA Pro}: Interactive Disassembler and Decompiler},
  year         = {2024},
  howpublished = {\url{https://hex-rays.com/ida-pro}},
  note         = {Accessed: June 2026}
}

@misc{gemini25,
  author        = {{Gemini Team, Google DeepMind}},
  title         = {Gemini 2.5: Pushing the Frontier with Advanced Reasoning, Multimodality, Long Context, and Next Generation Agentic Capabilities},
  year          = {2025},
  eprint        = {2507.06261},
  archivePrefix = {arXiv},
  primaryClass  = {cs.CL},
  url           = {https://arxiv.org/abs/2507.06261}
}

@misc{gpt54mini,
  author       = {{OpenAI}},
  title        = {Introducing {GPT-5.4} mini and nano},
  year         = {2026},
  howpublished = {\url{https://openai.com/index/introducing-gpt-5-4-mini-and-nano/}},
  note         = {Large language model; accessed June 2026}
}

@misc{haiku45,
  author       = {{Anthropic}},
  title        = {Introducing Claude Haiku 4.5},
  year         = {2025},
  howpublished = {\url{https://www.anthropic.com/news/claude-haiku-4-5}},
  note         = {Large language model (\texttt{claude-haiku-4-5}); accessed June 2026}
}

@misc{qwen3,
  title         = {Qwen3 Technical Report},
  author        = {{Qwen Team}},
  year          = {2025},
  eprint        = {2505.09388},
  archivePrefix = {arXiv},
  primaryClass  = {cs.CL},
  url           = {https://arxiv.org/abs/2505.09388}
}

@inproceedings{wang2022selfconsistency,
  title     = {Self-Consistency Improves Chain of Thought Reasoning in Language Models},
  author    = {Wang, Xuezhi and Wei, Jason and Schuurmans, Dale and Le, Quoc and Chi, Ed and Narang, Sharan and Chowdhery, Aakanksha and Zhou, Denny},
  booktitle = {International Conference on Learning Representations (ICLR)},
  year      = {2023},
  url       = {https://arxiv.org/abs/2203.11171}
}

@misc{du2023multiagentdebate,
  title         = {Improving Factuality and Reasoning in Language Models through Multiagent Debate},
  author        = {Du, Yilun and Li, Shuang and Torralba, Antonio and Tenenbaum, Joshua B. and Mordatch, Igor},
  year          = {2023},
  eprint        = {2305.14325},
  archivePrefix = {arXiv},
  primaryClass  = {cs.CL},
  url           = {https://arxiv.org/abs/2305.14325}
}

@misc{wang2024moa,
  title         = {Mixture-of-Agents Enhances Large Language Model Capabilities},
  author        = {Wang, Junlin and Wang, Jue and Athiwaratkun, Ben and Zhang, Ce and Zou, James},
  year          = {2024},
  eprint        = {2406.04692},
  archivePrefix = {arXiv},
  primaryClass  = {cs.CL},
  url           = {https://arxiv.org/abs/2406.04692}
}

@article{li2024moreagents,
  title   = {More Agents Is All You Need},
  author  = {Li, Junyou and Zhang, Qin and Yu, Yangbin and Fu, Qiang and Ye, Deheng},
  journal = {Transactions on Machine Learning Research (TMLR)},
  year    = {2024},
  url     = {https://arxiv.org/abs/2402.05120}
}

@misc{chen2025llmensemble,
  title         = {Harnessing Multiple Large Language Models: A Survey on {LLM} Ensemble},
  author        = {Chen, Zhijun and Li, Jingzheng and Chen, Pengpeng and Li, Zhuoran and Sun, Kai and Luo, Yuankai and Mao, Qianren and Yang, Dingqi and Sun, Hailong and Yu, Philip S.},
  year          = {2025},
  eprint        = {2502.18036},
  archivePrefix = {arXiv},
  primaryClass  = {cs.CL},
  url           = {https://arxiv.org/abs/2502.18036}
}

@misc{tsfaty2022malicious,
  title         = {Malicious Source Code Detection Using Transformer},
  author        = {Tsfaty, Chen and Fire, Michael},
  year          = {2022},
  eprint        = {2209.07957},
  archivePrefix = {arXiv},
  primaryClass  = {cs.CR},
  url           = {https://arxiv.org/abs/2209.07957}
}

@article{zhao2025apppoet,
  title   = {{AppPoet}: Large Language Model based Android Malware Detection via Multi-view Prompt Engineering},
  author  = {Zhao, Wenxiang and Wu, Juntao and Meng, Zhaoyi},
  journal = {Expert Systems with Applications},
  volume  = {262},
  pages   = {125546},
  year    = {2025},
  url     = {https://arxiv.org/abs/2404.18816}
}

@misc{qian2025lamd,
  title         = {{LAMD}: Context-driven Android Malware Detection and Classification with {LLMs}},
  author        = {Qian, Xingzhi and Zheng, Xinran and He, Yiling and Yang, Shuo and Cavallaro, Lorenzo},
  year          = {2025},
  eprint        = {2502.13055},
  archivePrefix = {arXiv},
  primaryClass  = {cs.CR},
  url           = {https://arxiv.org/abs/2502.13055}
}

@inproceedings{fujii2024feasibility,
  title     = {Feasibility Study for Supporting Static Malware Analysis Using {LLM}},
  author    = {Fujii, Shota and Yamagishi, Rei},
  booktitle = {Computer Security. ESORICS 2024 International Workshops},
  series    = {Lecture Notes in Computer Science},
  volume    = {15264},
  publisher = {Springer},
  year      = {2025},
  url       = {https://arxiv.org/abs/2411.14905}
}

@article{patsakis2024deobfuscation,
  title   = {Assessing {LLMs} in Malicious Code Deobfuscation of Real-world Malware Campaigns},
  author  = {Patsakis, Constantinos and Casino, Fran and Lykousas, Nikolaos},
  journal = {Expert Systems with Applications},
  volume  = {256},
  pages   = {124912},
  year    = {2024},
  doi     = {10.1016/j.eswa.2024.124912},
  url     = {https://arxiv.org/abs/2404.19715}
}

@misc{llama3,
  title         = {The {Llama} 3 Herd of Models},
  author        = {{Llama Team, AI @ Meta}},
  year          = {2024},
  eprint        = {2407.21783},
  archivePrefix = {arXiv},
  primaryClass  = {cs.AI},
  url           = {https://arxiv.org/abs/2407.21783}
}

@inproceedings{cao2022neurdp,
  title     = {Boosting Neural Networks to Decompile Optimized Binaries},
  author    = {Cao, Ying and Liang, Ruigang and Chen, Kai and Hu, Peiwei},
  booktitle = {Proceedings of the 38th Annual Computer Security Applications Conference (ACSAC)},
  year      = {2022},
  url       = {https://arxiv.org/abs/2301.00969}
}

@misc{dozono2024secure,
  title         = {Large Language Models for Secure Code Assessment: A Multi-Language Empirical Study},
  author        = {Dozono, Kohei and Espinha Gasiba, Tiago and Stocco, Andrea},
  year          = {2024},
  eprint        = {2408.06428},
  archivePrefix = {arXiv},
  primaryClass  = {cs.SE},
  url           = {https://arxiv.org/abs/2408.06428}
}

@misc{jiang2024cvd,
  title         = {Investigating Large Language Models for Code Vulnerability Detection: An Experimental Study},
  author        = {Jiang, Xuefeng and Wu, Lvhua and Sun, Sheng and Li, Jia and Xue, Jingjing and Wang, Yuwei and Wu, Tingting and Liu, Min},
  year          = {2024},
  eprint        = {2412.18260},
  archivePrefix = {arXiv},
  primaryClass  = {cs.SE},
  url           = {https://arxiv.org/abs/2412.18260}
}

@inproceedings{xia2023apr,
  title     = {Automated Program Repair in the Era of Large Pre-trained Language Models},
  author    = {Xia, Chunqiu Steven and Wei, Yuxiang and Zhang, Lingming},
  booktitle = {Proceedings of the 45th International Conference on Software Engineering (ICSE)},
  pages     = {1482--1494},
  year      = {2023},
  doi       = {10.1109/ICSE48619.2023.00129}
}

@article{zhang2024cybersecurity,
  title   = {When {LLMs} Meet Cybersecurity: A Systematic Literature Review},
  author  = {Zhang, J. and Bu, H. and Wen, H. and Chen, Y. and Li, L. and Zhu, H.},
  journal = {Cybersecurity},
  year    = {2025},
  doi     = {10.1186/s42400-025-00361-w},
  url     = {https://arxiv.org/abs/2405.03644}
}

@misc{xu2024llmcyber,
  title         = {Large Language Models for Cyber Security: A Systematic Literature Review},
  author        = {Xu, H. and Wang, S. and Li, N. and Wang, K. and Zhao, Y. and Chen, K. and Yu, T. and Liu, Y. and Wang, H.},
  year          = {2024},
  eprint        = {2405.04760},
  archivePrefix = {arXiv},
  primaryClass  = {cs.CR},
  url           = {https://arxiv.org/abs/2405.04760}
}

@misc{walton2025android,
  title         = {Exploring Large Language Models for Semantic Analysis and Categorization of Android Malware},
  author        = {Walton, Brandon J. and Khatun, Mst Eshita and Ghawaly, James M. and Ali-Gombe, Aisha},
  year          = {2025},
  eprint        = {2501.04848},
  archivePrefix = {arXiv},
  primaryClass  = {cs.CR},
  url           = {https://arxiv.org/abs/2501.04848}
}

@misc{lu2024malsight,
  title         = {{MalSight}: Exploring Malicious Source Code and Benign Pseudocode for Iterative Binary Malware Summarization},
  author        = {Lu, Haolang and Peng, Hongrui and Nan, Guoshun and Cui, Jiaoyang and Wang, Cheng and Jin, Weifei and Wang, Songlin and Pan, Shengli and Tao, Xiaofeng},
  year          = {2024},
  eprint        = {2406.18379},
  archivePrefix = {arXiv},
  primaryClass  = {cs.CR},
  url           = {https://arxiv.org/abs/2406.18379}
}

@misc{alkaraki2024malware,
  title         = {Exploring {LLMs} for Malware Detection: Review, Framework Design, and Countermeasure Approaches},
  author        = {Al-Karaki, Jamal and Khan, Muhammad Al-Zafar and Omar, Marwan},
  year          = {2024},
  eprint        = {2409.07587},
  archivePrefix = {arXiv},
  primaryClass  = {cs.CR},
  url           = {https://arxiv.org/abs/2409.07587}
}

\newpage
\appendix

\newpage
~\newpage
\section{Popular Decompilers}

Table~\ref{tab:decompilers} presents four of the most popular open-source and proprietary decompilers used in industry.

\begin{table}[htb!]
\centering
\footnotesize
\setlength{\tabcolsep}{4pt}
\renewcommand{\arraystretch}{1.08}
\begin{tabularx}{\columnwidth}{@{}>{\raggedright\arraybackslash}p{2.3cm}X@{}}
\toprule
\textbf{Decompiler} & \textbf{Availability / Cost} \\
\midrule
\textbf{Ghidra}
& Free, open source (NSA) \\
\addlinespace[2pt]

\textbf{RetDec}
& Free, open source (Avast) \\
\addlinespace[2pt]

IDA Pro (Hex-Rays)
& Paid, proprietary; free \mbox{IDA Free} tier for evaluation \\
\addlinespace[2pt]

Binary Ninja (Vector 35)
& Paid, proprietary; free cloud tier and Free edition for evaluation/education \\
\bottomrule
\end{tabularx}
\caption{Representative C decompilers and their availability. Tool-specific naming and type-recovery heuristics can produce divergent renderings of the same function. We focus on \textbf{Ghidra} and \textbf{RetDec}, both free and open source.}
\label{tab:decompilers}
\end{table}

\section{Metadata and size statistics}
\paragraph{Metadata and size statistics.}
For each sample, we maintain a metadata record containing the sample
identifier, ground-truth label, behavioral annotations, provenance
fields, and paths to the source, compiled object, Ghidra output, and
RetDec output. We also record basic size statistics for the compiled and
decompiled artefacts. Table~\ref{tab:dataset-stats} reports median sizes
and ranges by class. Although compiled object sizes differ across
classes, the decompiled inputs seen by the LLM are of comparable scale
for benign and malicious samples, reducing the risk that the task is
dominated by obvious length cues.

\begin{table}[htb!]
\centering
\small
\setlength{\tabcolsep}{4pt}
\begin{tabular}{@{}lcc@{}}
\toprule
Metric & Malicious & Benign \\
\midrule
Source lines        & 261 (20--2030)        & --- \\
Compiled (bytes)    & 6{,}128 (1.3k--30k)   & 16{,}468 (16.0k--17.0k) \\
Ghidra (bytes)      & 12{,}705 (1.5k--280k) & 10{,}802 (7.2k--69k) \\
RetDec (bytes)      & 14{,}374 (0.4k--142k) & 9{,}186 (5.2k--54k) \\
\bottomrule
\end{tabular}
\caption{Size statistics by class: median and min--max range. Decompiled
output sizes are comparable across benign and malicious samples for both
decompilers.}
\label{tab:dataset-stats}
\end{table}

\section{Gemini-Pro Reference Results}
\label{app:gemini-reference}

The main experiments focus on smaller or lower-cost models, since
industrial malware triage may require running inference over large
volumes of binaries and decompiler views. For reference, we also evaluate
\texttt{Gemini 2.5 Pro}, a larger model from the same family as
\texttt{Gemini 2.5 Flash-Lite}. Table~\ref{tab:gemini-reference} shows
that the larger model performs strongly from a single decompiler view,
with RetDec alone reaching the best $F_1$. However, multi-view prompting
substantially closes the gap between the lower-cost and larger Gemini
models: \texttt{Flash-Lite} improves from $75.0$ $F_1$ with its best
single view to $88.9$ with both views, reducing the gap to
\texttt{Gemini Pro}'s best result from $15.3$ points to $1.4$ points.
This supports the main motivation of our deployment setting: combining
decompiler views can recover much of the performance otherwise associated
with a larger model, while remaining more practical for large-scale
triage.

\begin{table}[htb!]
\centering
\small
\setlength{\tabcolsep}{7pt}
\begin{tabular}{lcc}
\toprule
\textbf{Setting} &
\texttt{Gemini Flash-Lite} &
\texttt{Gemini Pro} \\
\midrule
Ghidra only &
$75.0_{\scriptscriptstyle \pm 2.4}$ &
$83.7_{\scriptscriptstyle \pm 0.7}$ \\

RetDec only &
$73.4_{\scriptscriptstyle \pm 3.1}$ &
$\mathbf{90.3}_{\scriptscriptstyle \pm 1.4}$ \\

Multi-view &
$\mathbf{88.9}_{\scriptscriptstyle \pm 2.0}$ &
$85.4_{\scriptscriptstyle \pm 0.8}$ \\

Consensus &
$81.0$ &
$86.7$ \\
\bottomrule
\end{tabular}
\caption{Malicious-class $F_1$ (\%) for Gemini models. Subscripts denote
standard deviation over five runs where applicable. \texttt{Gemini Pro}
is included as a larger-model reference point. Multi-view prompting lets
the lower-cost \texttt{Flash-Lite} model recover most of the gap to
\texttt{Gemini Pro}'s best single-view performance.}
\label{tab:gemini-reference}
\end{table}

\section{Per-Family Recall Breakdown}
\label{app:category-breakdown}
Recall on each malware family (Ghidra / RetDec / Combined), per model,
expanding the aggregate recall trends from Section~\ref{sec:experiments}.
Family sizes ($n$) are given on the $x$-axis; single-example families
($n{=}1$) are high-variance and shown for completeness only.

\subsection{Per-family malware recall}
Recall on each malware family (Ghidra / RetDec / Combined), per model.
Family sizes ($n$) are given on the $x$-axis; single-example families
($n{=}1$) are high-variance and shown for completeness only.
 
\begin{figure*}[p]
\centering
\includegraphics[width=\textwidth]{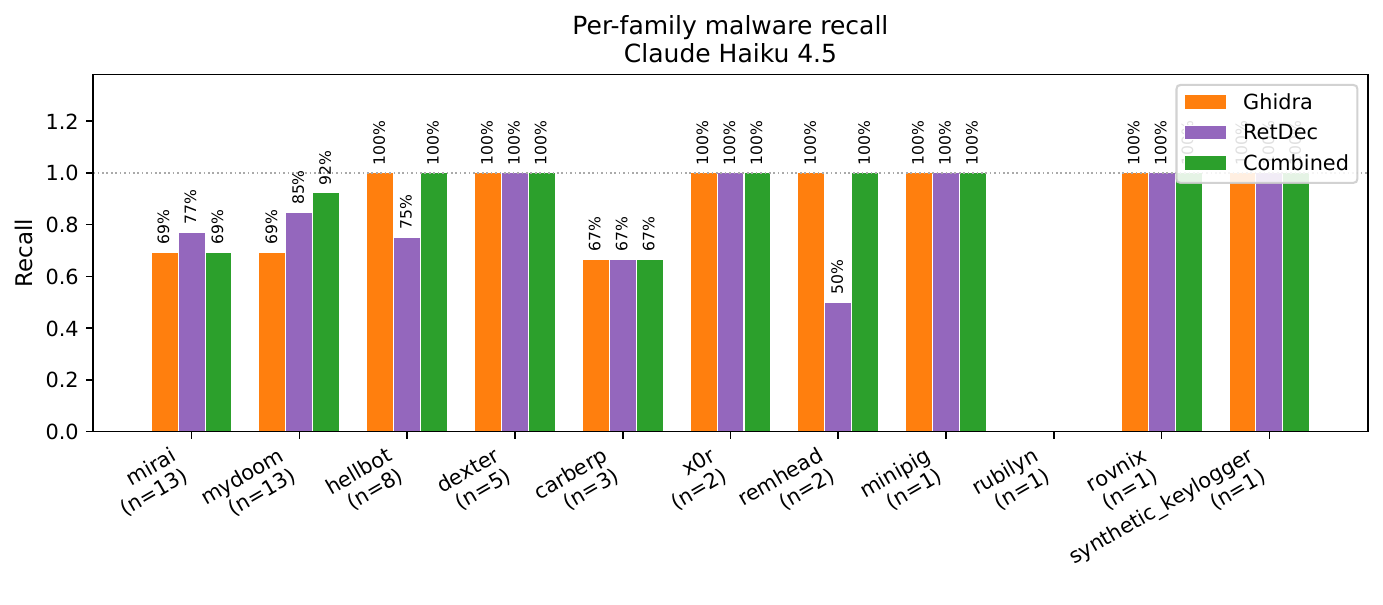}
\caption{Per-family recall --- Claude Haiku 4.5.}
\label{fig:app-family-haiku}
\end{figure*}
 
\begin{figure*}[p]
\centering
\includegraphics[width=\textwidth]{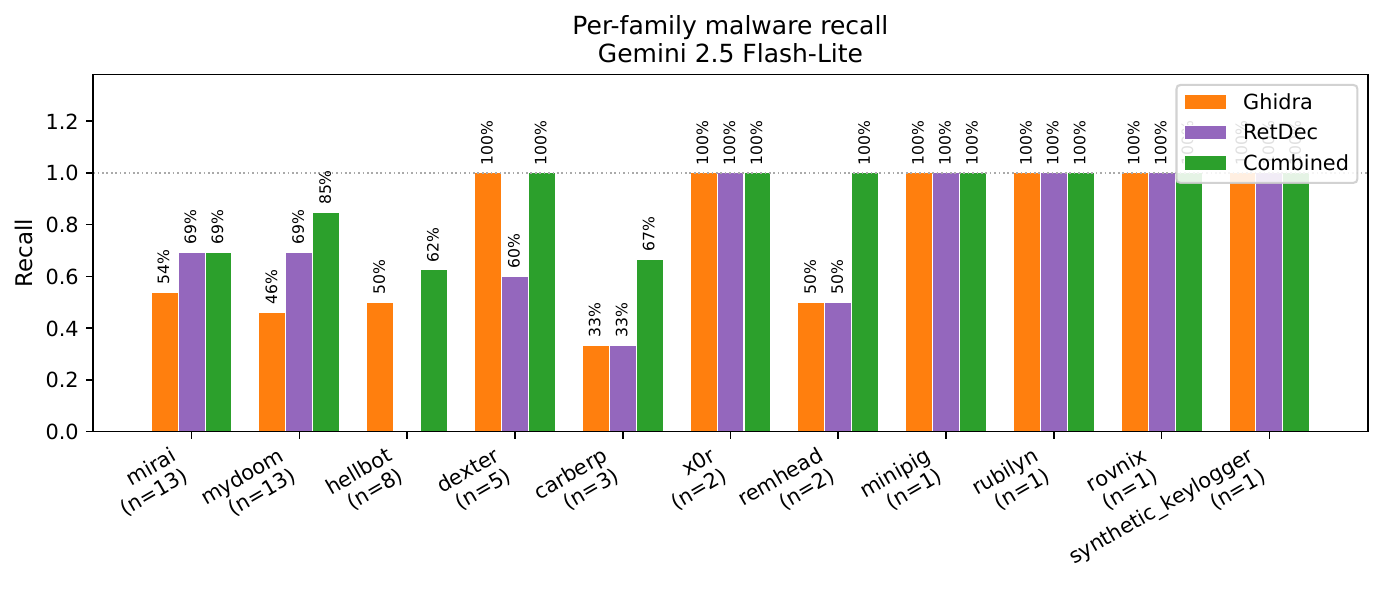}
\caption{Per-family recall --- Gemini 2.5 Flash-Lite.}
\label{fig:app-family-flashlite}
\end{figure*}
 
\begin{figure*}[p]
\centering
\includegraphics[width=\textwidth]{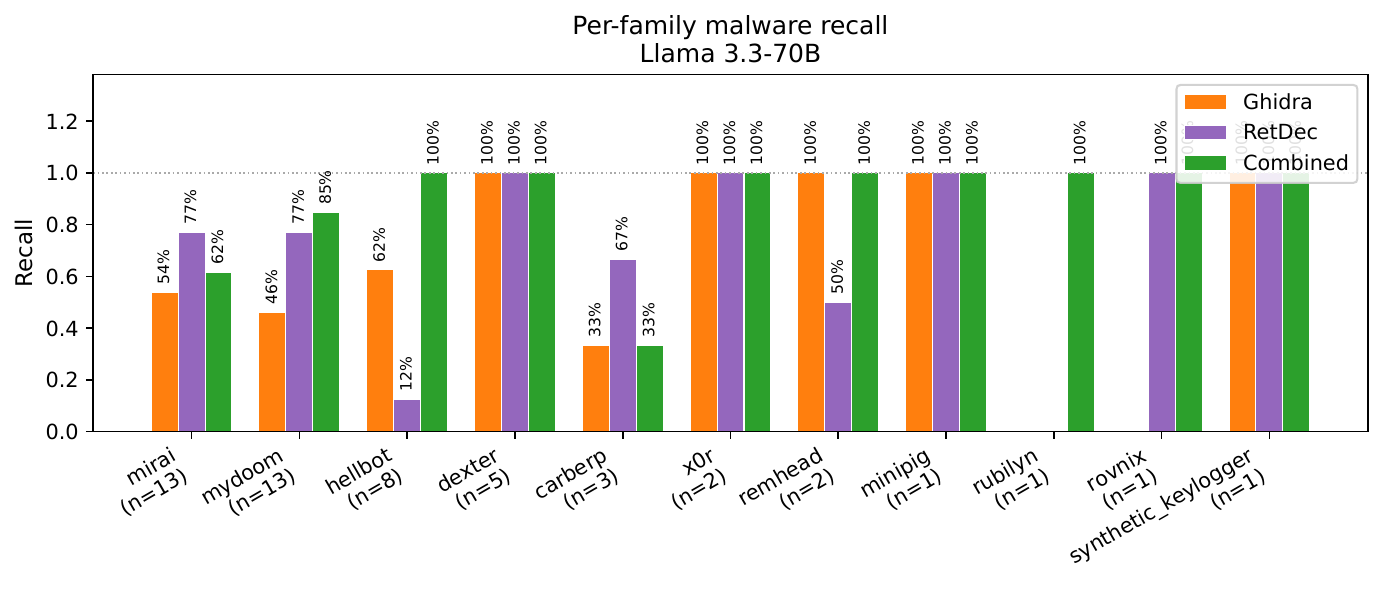}
\caption{Per-family recall --- Llama-3.3-70B-Instruct.}
\label{fig:app-family-llama}
\end{figure*}
 
\begin{figure*}[p]
\centering
\includegraphics[width=\textwidth]{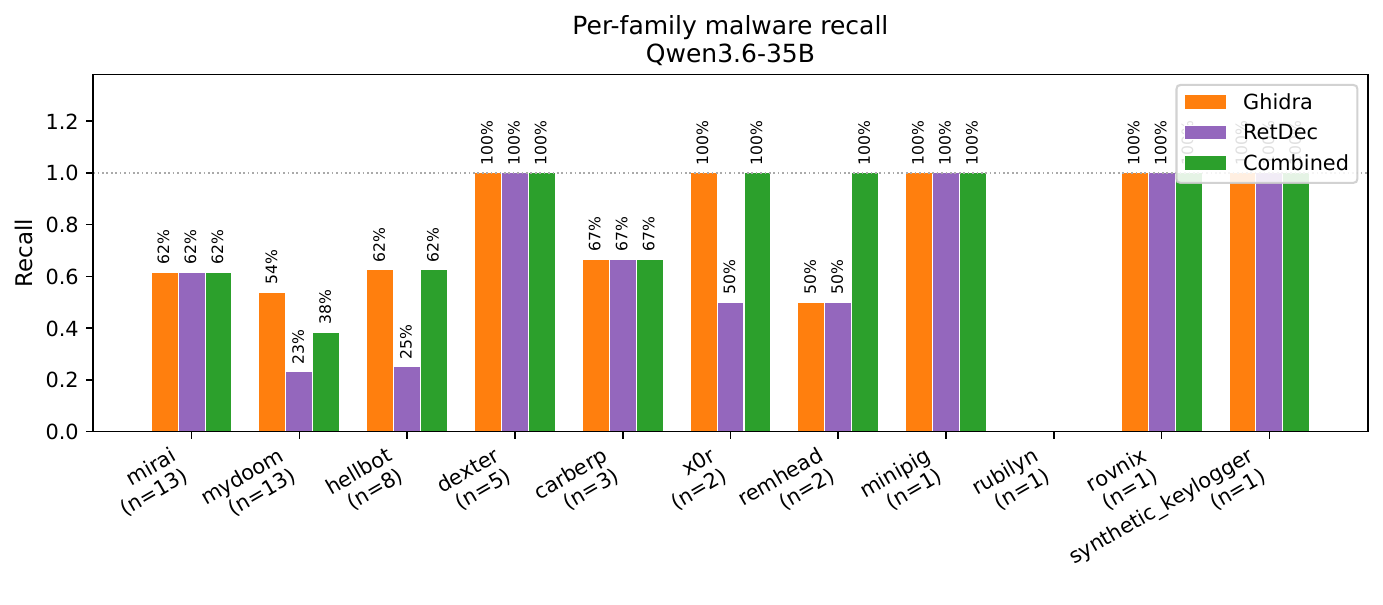}
\caption{Per-family recall --- Qwen3.6-35B-A3B.}
\label{fig:app-family-qwen}
\end{figure*}
 
\begin{figure*}[p]
\centering
\includegraphics[width=\textwidth]{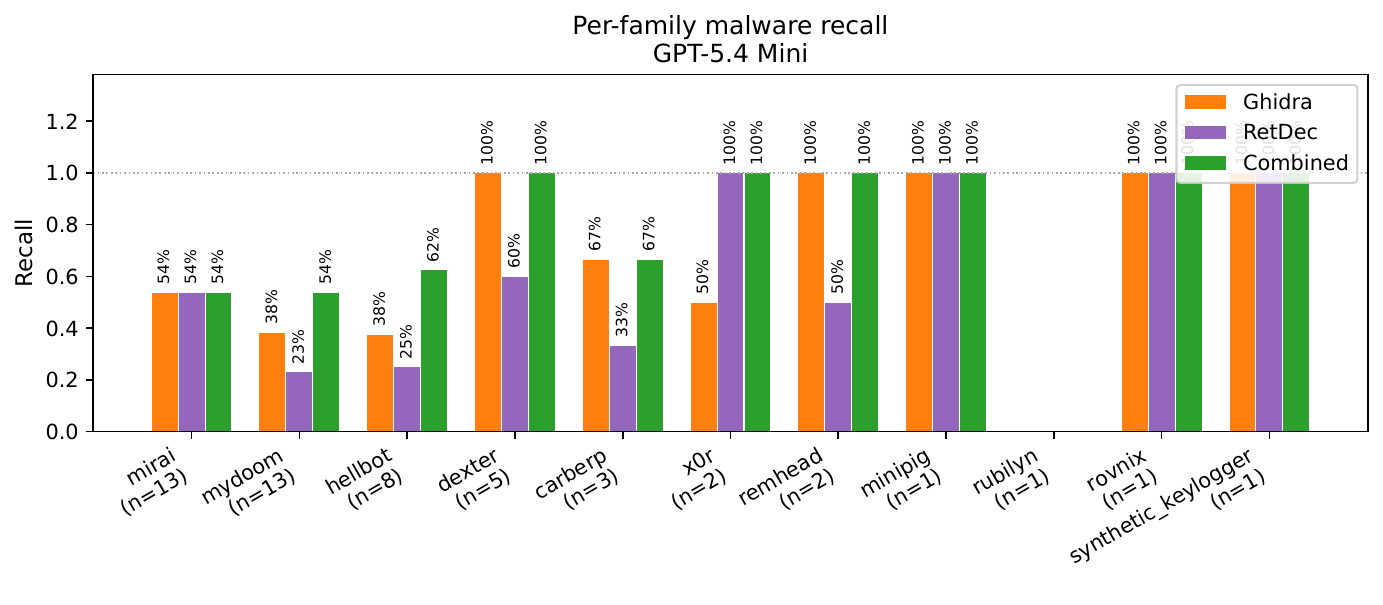}
\caption{Per-family recall --- GPT-5.4-mini.}
\label{fig:app-family-gpt}
\end{figure*}

\begin{figure*}[t]
\centering

\begin{subfigure}{0.32\textwidth}
\centering
\includegraphics[width=\linewidth]{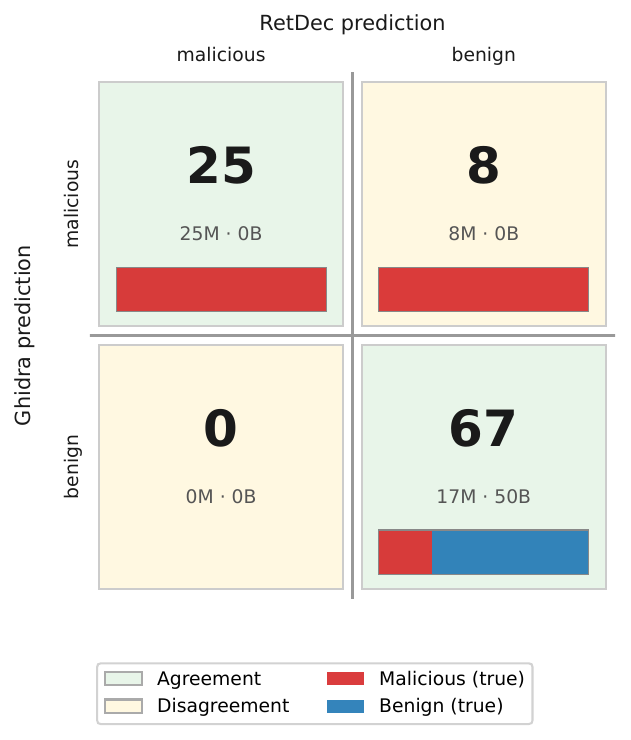}
\caption{\texttt{Qwen3}}
\end{subfigure}
\hfill
\begin{subfigure}{0.32\textwidth}
\centering
\includegraphics[width=\linewidth]{latex/figures/fig_disagreement_matrix_meta-llama-Llama-3.3-70B-Instruct-Turbo.pdf}
\caption{\texttt{Llama}}
\end{subfigure}
\hfill
\begin{subfigure}{0.32\textwidth}
\centering
\includegraphics[width=\linewidth]{latex/figures/fig_disagreement_matrix_claude-haiku-4-5.pdf}
\caption{\texttt{Haiku}}
\end{subfigure}
\hfill
\begin{subfigure}{0.32\textwidth}
\centering
\includegraphics[width=\linewidth]{latex/figures/fig_disagreement_matrix_gemini-2.5-flash-lite.pdf}
\caption{\texttt{Flash-Lite}}
\end{subfigure}
\hfill
\begin{subfigure}{0.32\textwidth}
\centering
\includegraphics[width=\linewidth]{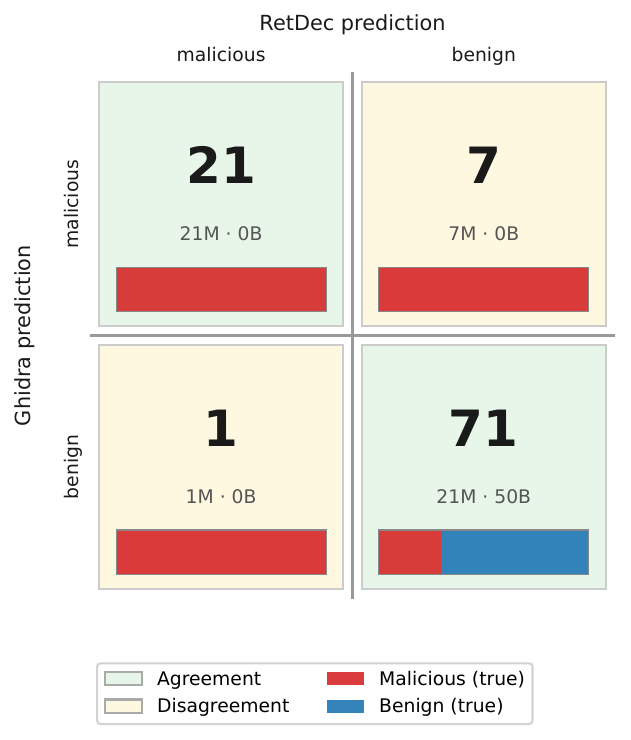}
\caption{\texttt{GPT-mini}}
\end{subfigure}

\caption{Prediction agreement between Ghidra-only and RetDec-only
classifiers. Rows correspond to the Ghidra prediction and columns to the
RetDec prediction. Off-diagonal cells show samples where the two
decompiler views lead to different labels.}
\label{fig:appendix-agreement}
\end{figure*}

\section{Prompt Templates}
\label{app:prompts}

\lstdefinestyle{prompt}{
  basicstyle=\ttfamily\small,
  breaklines=true,
  breakautoindent=true,
  columns=fullflexible,
  keepspaces=true,
  showstringspaces=false,
  frame=single,
  framesep=5pt,
  xleftmargin=6pt,
  xrightmargin=6pt,
  aboveskip=4pt,
  belowskip=4pt,
  literate={—}{{---}}1 {–}{{--}}1 {’}{{'}}1 {‘}{{`}}1 {“}{{``}}1 {”}{{''}}1
}

We use a definition-only prompt for the main experiments. It defines
malicious and benign behavior without listing concrete malware
indicators, to avoid steering the model toward particular attack families
or surface patterns. In the single-decompiler setting, the user message
contains one decompiled pseudo-C view. In the multi-view setting, the
same definition is used, but the user message contains both Ghidra and
RetDec outputs and states that they are complementary views of the same
binary; the exact prompt wrapper is provided in the released codebase. In
all settings, the model is asked to provide a brief rationale before
returning a structured binary decision.

\begin{figure*}[t]
\begin{lstlisting}[style=prompt]
You are a malware analyst specialising in static analysis of decompiled C code.
You will be given decompiled C source code recovered from a compiled binary.
Your task is to determine whether the code is malicious.

Malicious code is software that intentionally performs harmful, covert, or
unauthorised actions against a system or its users without their knowledge or
consent. It operates outside the user's awareness or against their interests,
regardless of how it achieves this.

Benign code performs legitimate, transparent, and well-scoped operations that
serve an obvious, non-harmful purpose and do not attempt to conceal their
behavior or effects.

Briefly explain what the code appears to do, then decide whether it is benign
or malicious.

Return your answer as JSON with the following fields:
{
  "rationale": "<brief explanation>",
  "decision": 0 or 1
}

Use decision = 1 for malicious code and decision = 0 for benign code.
\end{lstlisting}
\caption{Definition-only classification prompt used in the main
experiments. The decompiled code is supplied in the user message.}
\label{app:prompt-definition}
\end{figure*}

For the guided upper-bound experiment in Appendix~\ref{app:guided}, we
also use an indicator-guided prompt. This prompt augments the same
definition with an explicit checklist of suspicious behaviors.

\begin{figure*}[t]
\begin{lstlisting}[style=prompt]
You are a malware analyst specialising in static analysis of decompiled C code.
You will be given decompiled C source code recovered from a compiled binary.
Your task is to determine whether the code is malicious.

Malicious code is software that intentionally performs harmful, covert, or
unauthorised actions against a system or its users without their knowledge or
consent. Common indicators include:
- Network scanning, port sweeping, or raw socket abuse
- Remote shell or command execution
- Process injection or memory manipulation
- Keylogging, credential harvesting, or screen capture
- Persistence mechanisms
- DDoS or flood attack functions
- Botnet command-and-control communication
- Anti-debugging, anti-analysis, or sandbox evasion
- Encryption used to hide payloads or C2 traffic
- Destruction of files, logs, or system integrity

Benign code performs legitimate, transparent, and well-scoped operations that
serve an obvious, non-harmful purpose and do not attempt to conceal their
behavior or effects.

Briefly explain what the code appears to do, then decide whether it is benign
or malicious.

Return your answer as JSON with the following fields:
{
  "rationale": "<brief explanation>",
  "decision": 0 or 1
}

Use decision = 1 for malicious code and decision = 0 for benign code.
\end{lstlisting}
\caption{Indicator-guided prompt used only for the upper-bound analysis in
Appendix~\ref{app:guided}. It is not used in the main results.}
\label{app:prompt-guided}
\end{figure*}

\section{Guided-Prompt Upper Bound}
\label{app:guided}

The main experiments use the definition-only prompt in
Appendix~\ref{app:prompt-definition}. This prompt deliberately avoids
listing concrete malicious indicators, since such a list may steer the
model toward behaviors represented in the benchmark rather than testing
whether it can infer maliciousness from the decompiled code itself. To
estimate how much headroom a more guided prompt can provide, we also
evaluate the indicator-guided prompt in Appendix~\ref{app:prompt-guided}
on \texttt{Gemini 2.5 Flash-Lite}, the model with the largest multi-view
gain in the main results.

The guided prompt should be interpreted as an optimistic upper bound, not
as the main deployable setting. Its indicator list gives the model a
strong prior over suspicious behaviors, which can improve performance on
known malware patterns but may reduce robustness to novel behaviors or
benign programs with superficially similar APIs. We therefore report it
only as an ablation.

\begin{table}[h]
\centering
\small
\begin{tabular}{lcc}
\toprule
\textbf{Setting} & \textbf{Definition-only} & \textbf{Indicator-guided} \\
\midrule
Ghidra only & 75.0 & 88.9 \\
RetDec only & 73.4 & 90.1 \\
Multi-view  & 88.9 & 85.1 \\
Consensus   & 81.0 & 86.4 \\
\bottomrule
\end{tabular}
\caption{Malicious-class $F_1$ (\%) for \texttt{Gemini 2.5 Flash-Lite}
under the definition-only and indicator-guided prompts. The guided prompt
raises single-view performance to roughly the level of the definition-only
multi-view setting, indicating that explicit behavioral guidance can
recover much of the same signal. We treat this as an optimistic upper
bound rather than the main experimental setting.}
\label{tab:guided-upperbound}
\end{table}

\end{document}